\documentclass[%
 aps,
 apl,%
 amsmath,amssymb,
 reprint,%
]{revtex4-1}

\usepackage{bm}
\usepackage{natbib}
\usepackage{graphicx}
\graphicspath{ {images/} }

\begin{document}

\title[Graphene Transport Mediated by Micropatterned Substrates]{Graphene Transport Mediated by Micropatterned Substrates}

\author{J. Henry Hinnefeld}
 \affiliation{Department of Physics and Materials Research Laboratory, University of Illinois at Urbana-Champaign, 1110 West Green Street, Urbana, Illinois 61801, USA}
\author{Nadya Mason}
 \email{nadya@illinois.edu}
 \affiliation{Department of Physics and Materials Research Laboratory, University of Illinois at Urbana-Champaign, 1110 West Green Street, Urbana, Illinois 61801, USA}
 
\date{\today}

\begin{abstract}
Engineered substrates offer a promising avenue towards graphene devices having tunable properties.
In particular, topographically patterned substrates can expose unique behavior due to their ability
to induce local variations in strain and electrostatic doping.
However, to explore the range of possible science and applications, it is important to create topographic
substrates which both have tunable features and are suitable for transport measurements.
In this Letter we describe the fabrication of tunable, topographically patterned substrates suitable for transport measurements.
We report both optical and transport measurements of graphene devices fabricated on these substrates,
and demonstrate characteristic strain and local doping behavior induced by the topographic features.
\end{abstract}


\maketitle

Graphene is a material with enormous potential for both scientific research and technical applications
\cite{novoselov2004electric,novoselov2005two,zhang2005experimental,geim2007rise}.
In particular, the ability to tune graphene's properties through the use of engineered substrates offers a practical
method to explore graphene's properties and modify them for specific applications\cite{guinea2010energy, zhou2007substrate}.
Previous work on engineered substrates has employed substrate topography\cite{Tomori2011, mi2015creating, babichev2015influence, reserbat2014strain},
electrostatic charge injection\cite{chiu2010controllable},
substrate lattice mis-match\cite{zhou2007substrate},
and ferroelectric polarization\cite{hinnefeld2016single}
to achieve a range of modifications to graphene's properties.

\begin{figure}
\centering
\includegraphics[width=0.8\columnwidth]{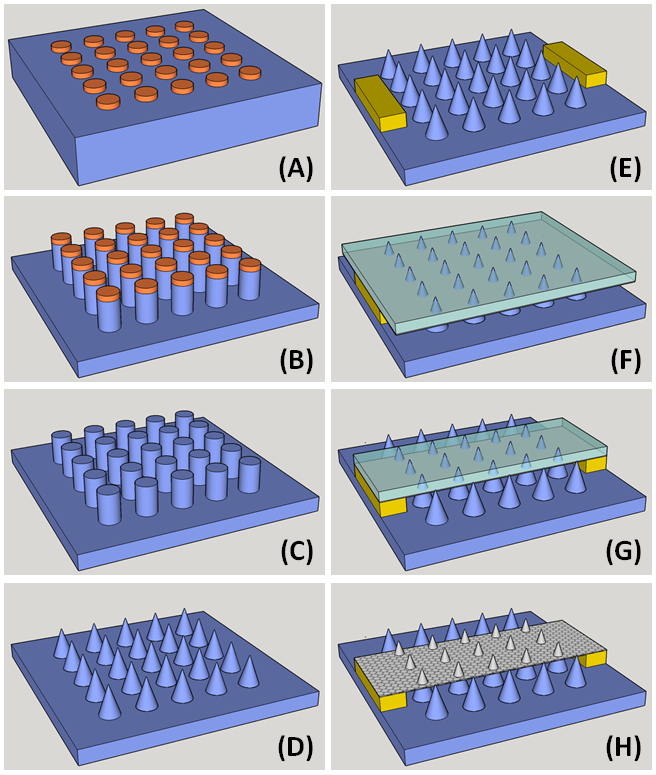}
\caption{Fabrication procedure for creating graphene devices on topographically patterned substrates (see text).}
\label{'fig:fab'}
\end{figure}

\begin{figure}
\centering
\includegraphics[width=\columnwidth]{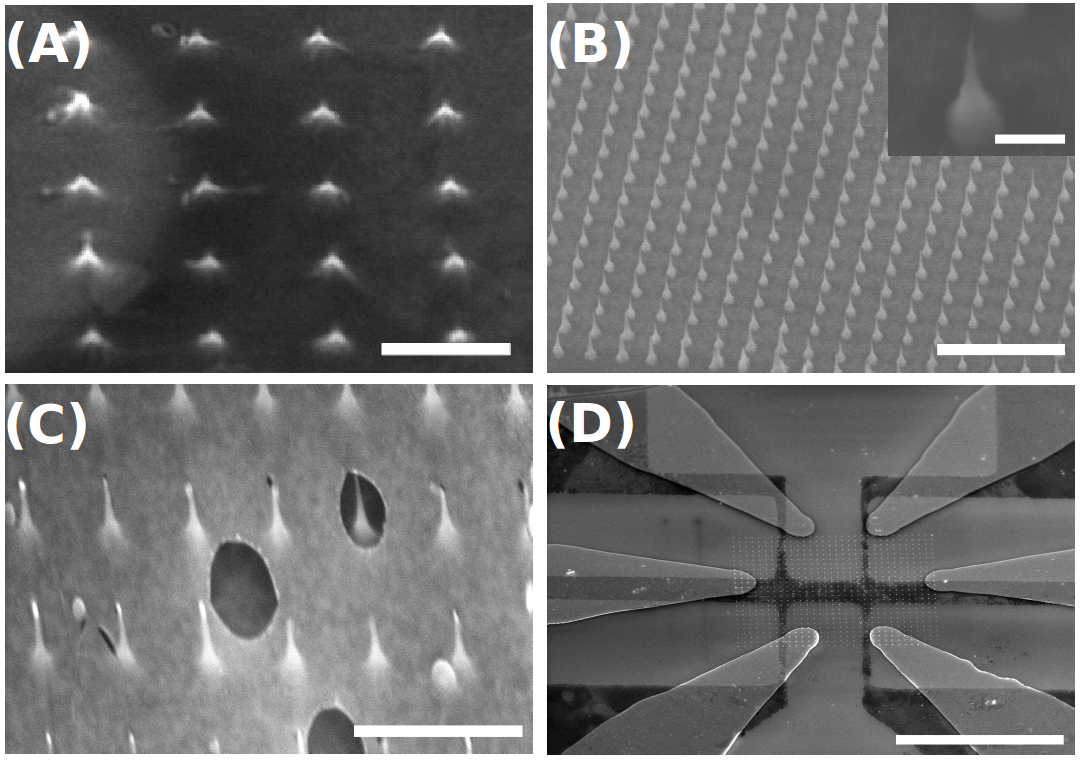}
\caption{SEM micrographs of substrates prepared by this method. 
(A) Graphene deposited on widely spaced topographic features partially delaminates in the vicinity of the SiO$_2$ cones. 
The lighter region on the left is an Au electrical lead.
The scale bar is 500 nm.
(B) After the BOE dip the SiO$_2$ pillars are sharpened into cones with a tip diameter of less than 20 nm.
The scale bar is 1 $\mu$m. Inset: A single sharpened cone. The scale bar is 100 nm.
(C) For tight topographic feature spacings the graphene is suspended on the pointed tips of the substrate features.
Here, a slightly ripped region of the graphene is used to show the substrate below.
The scale bar is 500 nm.
(D) After transfer the graphene is patterned in a Hall bar geometry. The six triangular features are Ti/Au
electrical leads. The scale bar is 40 $\mu$m.}
\label{'fig:sem'}
\end{figure}

Of the various substrate engineering techniques, topographic substrate patterning has two distinct advantages:
first, topographic substrates can create local strain in graphene. Strain has large effects on graphene's
electrical properties \cite{pereira2009strain}, 
from inducing minigaps \cite{ni2008uniaxial} 
to creating large pseudo-magnetic fields \cite{guinea2010energy, levy2010strain}.
To date however the techniques used to produce strain in graphene are either not 
amenable to performing electrical transport measurements on graphene
\cite{levy2010strain, Tomori2011, mohiuddin2009uniaxial, ni2008uniaxial, gill2015mechanical, reserbat2014strain}
or not compatible with standard lithographic fabrication procedures \cite{babichev2015influence}.
Second, topographic substrates can modulate the effect of a single electrostatic gate to produce 
complex doping profiles in graphene without the need for multiple, distinct gate electrodes.
Here we demonstrate a fabrication procedure for producing engineered arrays of topographic features on standard silicon
substrates, and we report optical measurements of strain and transport measurements of local doping in graphene
devices fabricated atop these substrates.


\begin{figure*}
\centering
\includegraphics[width=2\columnwidth]{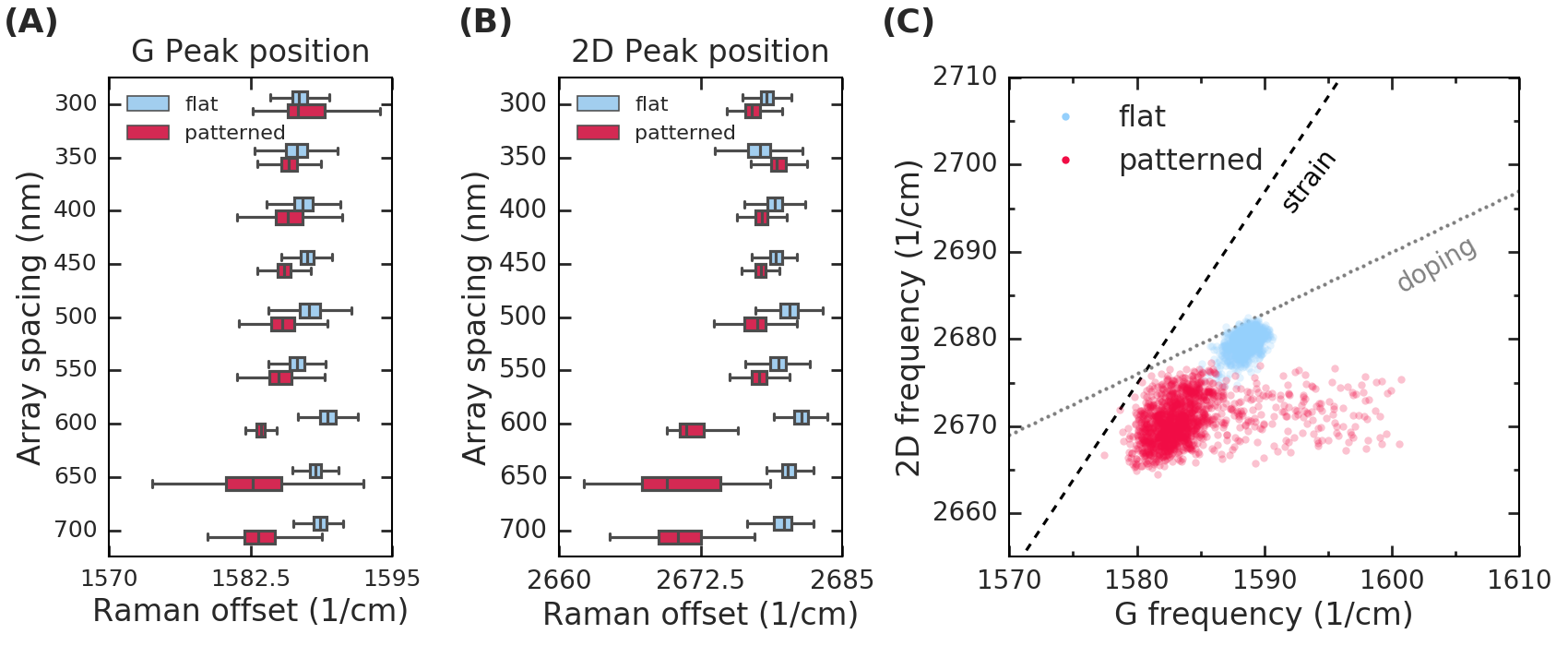}
\caption{Raman peak positions for graphene on micropatterned and flat substrates. 
    (A-B) Raman G and 2D peak positions extracted from a raster scan of graphene on
    topographically patterned and flat substrates for array spacings from 300 nm to 700 nm. 
    Each box element describes the distribution of measurements over all raster points in a given scan: 
    the central vertical line is the median, the colored box encompasses the interquartile range, and the horizontal lines
    denote the largest (smallest) non-outlier measurement.
    (C) Raman G peak position vs. 2D peak position for graphene on a flat substrate
    and a substrate with topographic features spaced 700 nm apart. The black dashed line shows the ratio 
    $r = \frac{\Delta \, \omega_\text{2D}}{\Delta \, \omega_\text{G}}$ expected for shifts due to
    strain, and the grey dotted line shows the ratio expected for shifts due to doping.
    The two lines intersect at the expected peak positions for undoped, unstrained graphene \cite{lee2012optical}.
    }
\label{'fig:raman'}
\end{figure*}

The process used to create the topographic features on the substrate is illustrated schematically in Figures \ref{'fig:fab'}A-D.
First an array of 20 nm thick copper circles is deposited, using standard electron-beam lithography and evaporation techniques, 
on a silicon chip covered with a 1000 nm layer of thermal oxide. 
The deposited copper is then used as a mask in a CF$_4$ reactive ion etching 
(RIE) step to produce cylindrical pillars in the SiO$_2$ layer. 
The RIE etch time and the diameter of the deposited copper mask circles together define the aspect ratio of the resulting pillars. 
For these devices we use a mask diameter of 100 nm, and a 10 minute etch time, which gives a height of approximately 200 nm.
Pillar diameter is independent of RIE etch time and is equal to the mask diameter.
After the etch, the copper mask is removed by immersing the chip 
in a 0.1M solution of ammonium persulfate for several hours. 
Finally, the chip is dipped in buffered oxide etchant 
to sharpen the SiO$_2$ pillars produced during the RIE step into pointed, conical shapes
having tips of {\raise.17ex\hbox{$\scriptstyle\sim$}20 nm diameter.

Graphene devices are fabricated on substrates prepared by this method using the process shown in Figures \ref{'fig:fab'}E-F. 
First Ti/Au (5 nm/30 nm) leads and contact pads are defined and deposited using electron-beam lithography and evaporation. 
Next, a monolayer of graphene is grown on a different substrate using established chemical vapor deposition techniques\cite{Li2009}. 
The graphene is then transferred to the topographic substrate using standard polymer-assisted wet-transfer 
techniques\cite{li2009transfer}. The same polymer layer used to transfer the graphene is then used as a 
resist in an electron-beam lithography step. Next the exposed graphene is removed using a reactive ion etch,
yielding graphene in a Hall bar configuration. Finally the remaining 
polymer resist is dissolved in acetone and the chip is dried in a critical point drying apparatus. 
Figure \ref{'fig:sem'} shows scanning electron microscope (SEM) micrographs of substrates and
graphene devices produced by this process.
Although the transfer and drying process does lead to some ripping of the graphene (see holes
evident in Figure \ref{'fig:sem'}C) holes and rips cover less than 10\% of the surface,
leaving the graphene largely robust for transport measurements on the 10 - 50 micron length scale of typical devices.
As shown in Figure \ref{'fig:sem'}A, we do not observe significant ripple formation \cite{babichev2015influence, reserbat2014strain}, likely
because our topographic features are widely spaced \cite{yamamoto2012princess}.


We perform optical measurements of graphene devices fabricated on these substrates 
to confirm the presence of strain. Raman spectroscopy iss performed using a Nanophoton 
Raman 11 microscope with a 532 nm laser at room temperature. The laser power iss kept below 1 mW to minimize local heating.
Raman measurements are collected in a raster pattern across a 
20 $\mu$m  $\times$ 20 $\mu$m area with a measurement spot size of 350 nm;
each scan encompasses a varying number of topographic
features depending on the array spacing. 
Random variations in spectra for regions away from topographic features yield shifts of less than $\pm$0.5 cm$^{-1}$.
The data for the flat substrates is collected separately for each device to account for
the varying residual doping present in each sample.
At each raster point the Raman G and 2D peak positions are extracted\cite{ferrari2006raman}. 
Figures \ref{'fig:raman'}A and \ref{'fig:raman'}B summarize the extracted positions 
of the Raman G and 2D peaks, respectively, for graphene on topographically patterned and flat regions of 
devices prepared by the method described above. Data is shown for devices having pillar spacings between 300 and 700 nm.

\begin{figure*}
\centering
\includegraphics[width=1.8\columnwidth]{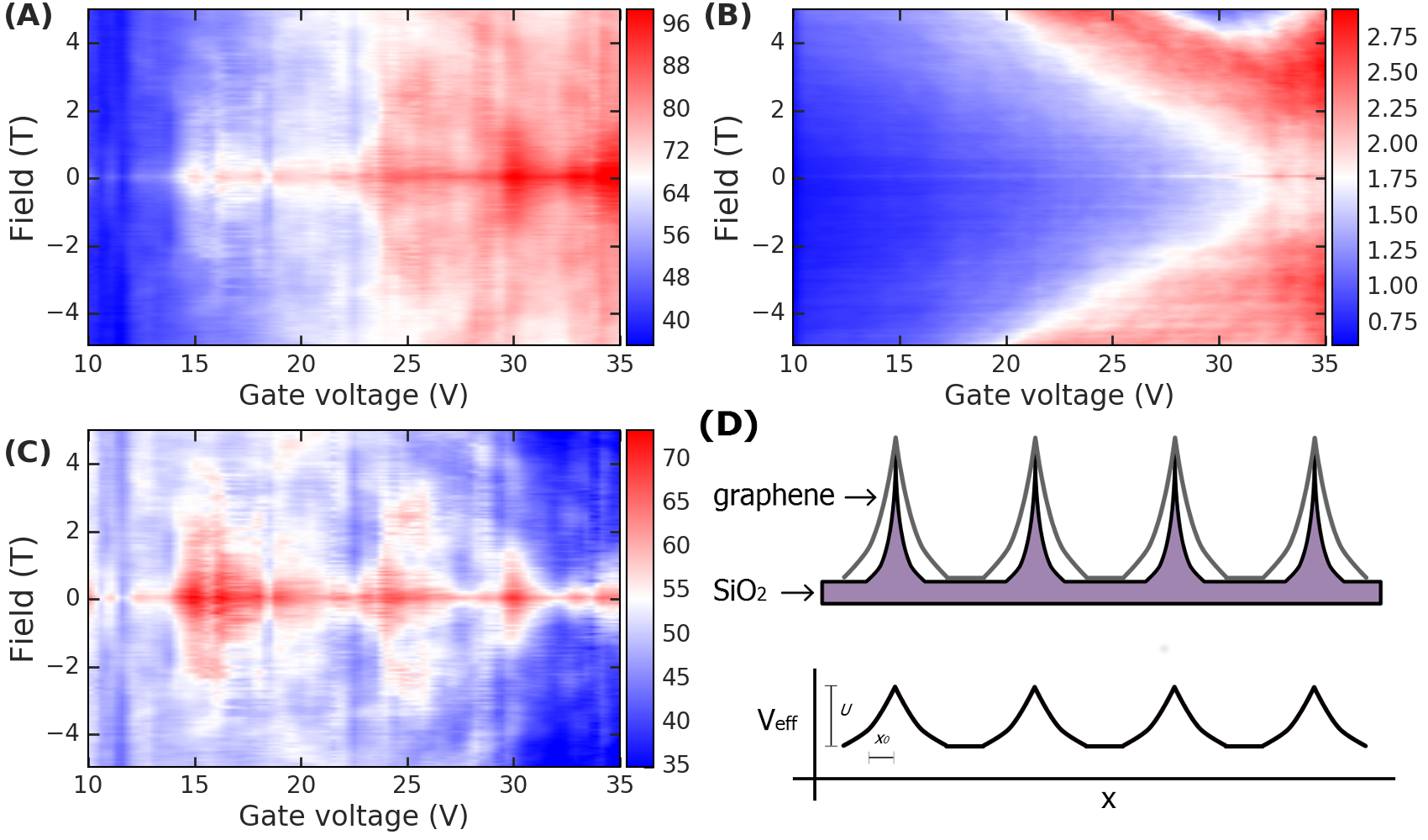}
\caption{
    Longitudinal resistance $R_\text{xx}$ of a graphene device fabricated on (A) a 
    substrate with 750 nm-spaced topographic features and (B) a flat substrate as a function of gate voltage and magnetic field. 
    The color scale is measured in k$\Omega$.
    (C) The same data as in (A) with a linear background $R_\text{background} = m V_g + b$ subtracted.
    (D) The potential profile created by the local delamination of the graphene. In the delamination
    regions the vacuum layer alters the gate capacitance, and thus creates local variations in the potential. 
}
\label{'fig:transport'}
\end{figure*}

Both the G and 2D peaks of graphene on the patterned 
substrate regions display shifted peak positions relative to graphene on the flat regions. 
This shift increases with increasing topographic feature spacing and displays a qualitative jump
for spacings above 600 nm. We attribute this jump to a 
snap-through transition\cite{gill2015mechanical, scharfenberg2012observation}
in the adhesion of the graphene to the substrate: for spacings below 600 nm the graphene is suspended
in the entire topographically patterned region, while for spacings above 600 nm the graphene adheres to the substrate
except in the immediate vicinity of a topographic feature. The partial delamination
present in the sparse topographic samples
produces strain in the graphene which generates the shifted Raman peak positions \cite{gill2015mechanical}.

Doping from charge impurities in the substrate is also known to 
shift Raman peak positions in graphene\cite{reina2008large, casiraghi2007raman}, however the ratio 
\begin{equation}
    r = \frac{\Delta \, \omega_\text{2D}}{\Delta \, \omega_\text{G}}
\end{equation}
(where $\Delta \, \omega$ is the shift in a Raman peak position relative to its intrinsic value)
differs between the two mechanisms \cite{lee2012optical}. 
Experimental measurements \cite{zabel2012raman, metzger2009biaxial, ding2010stretchable} 
and theoretical results\cite{mohr2010splitting, mohiuddin2009uniaxial} place the ratio for strain 
between 2.25 and 2.8 and the ratio for doping at approximately 0.75 \cite{lee2012optical}. 
Figure \ref{'fig:raman'}C shows the extracted Raman G and 2D peak positions for a representative
topographically patterned sample having a spacing of 700 nm,
along with lines corresponding to $r_\text{strain}$ (dashed) and $r_\text{doping}$ (dotted). 
The difference between the topographic and flat graphene samples lies along $r_\text{strain}$; thus, we
attribute the shifted peak positions to the effect of strain in the graphene.

Next we perform magneto-transport measurements on a device having a substrate pillar spacing of 750 nm,
to elucidate the effect of local variations in the electrostatic potential. Several similar devices were
measured and yielded qualitatively similar results.
For sparsely patterned substrates the graphene is locally delaminated in the vicinity of 
each individual topographic feature. This delamination alters the 
effective gate capacitance by including a region of vacuum in series with the SiO$_2$ dielectric layer.
The local variation in the gate capacitance creates a corresponding variation in the potential,
effectively creating a circular potential barrier, e.g. a quantum dot.
This situation is illustrated schematically in Figure \ref{'fig:transport'}D.
Carriers in graphene cannot be confined electrostatically: as massless particles governed by the Dirac equation
(in the low energy limit)
they display Klein tunneling \cite{katsnelson2006chiral,young2009quantum}.
However, previous work has shown that circular potential barriers can create pseudo-bound states in 
graphene \cite{silvestrov2007quantum,matulis2008quasibound,hewageegana2008electron,chen2007fock, heinisch2013mie}.
For graphene on topographically patterned substrates we therefore expect transport behavior to display signatures
of scattering from these pseudo-bound states.
We note that we do not expect coherent transport behavior (such as Fabry-P\'erot resonance or superlattice 
effects) across multiple topographic features, as the coherence length is {\raise.17ex\hbox{$\scriptstyle\sim$}1 micron \cite{berger2006electronic}), 
and a typical device length is 20 $\mu$m which encompasses {\raise.17ex\hbox{$\scriptstyle\sim$}30 pillars.


Figures \ref{'fig:transport'}A and \ref{'fig:transport'}B show the results of magneto-transport 
measurements performed at 250 mK on a 750 nm-spaced topographic device and a flat control device, respectively.
Both devices are 20 $\mu$m long and 10 $\mu$m wide; for the topographic device the patterned substrate features cover the
entire device area.
The Dirac point for both devices is located at approximately 40 V; this reflects both the residual doping
from the fabrication process as well as the reduced gate capacitance of the thicker-than-normal SiO$_2$ dielectric layer.
All transport measurements were taken with a channel current of 50 nA, and gate leakage current was always less than five percent
of channel current. The reduced mobility of the topographic device relative to the flat control device is due
to rips introduced during the critical point drying step of the fabrication process.
Qualitative differences between the topographic and flat devices are apparent: 
the flat control device displays the onset of a typical Landau level fan pattern\cite{bolotin2009observation}, 
however the topographic device displays several resistance maxima not present in the control device.
Figure \ref{'fig:transport'}C shows the same data as Figure \ref{'fig:transport'}A with a linear background 
$R_\text{background} = m V_g + b$ subtracted; we remove this background to emphasize deviations from the
expected Dirac cone pattern of gated graphene. Several additional local maxima are visible in the low gate voltage region.
The diamond-like high resistance features (red regions in Figure \ref{'fig:transport'}C) are reproducible.
Although they are somewhat irregular, they have typical energy scales of 
{\raise.17ex\hbox{$\scriptstyle\sim$}}1 T in magnetic field,
and {\raise.17ex\hbox{$\scriptstyle\sim$}}5 V in gate voltage.

The scale of the observed transport features, in both magnetic field and gate voltage,
is in good agreement with theoretical predictions for scattering from 
quasi-bound states in graphene.
Considering first the magnetic field scale,
previous work\cite{heinisch2013mie} has shown that the scattering properties of
quantum dots in graphene depend on the size of the dot:
for small dots forward scattering is strongly suppressed and conductivity is reduced, while large dots can focus
carriers and enhance conductivity. 
In the presence of a magnetic field, we take the cyclotron radius $r_c$ to be the relevant length scale,
e.g. for dot size $r < r_c$ we expect an effective ``small'' dot with reduced conductance,
while for $r > r_c$ we expect a ``large'' dot with larger conductance. This is consistent with our observation
of more resistive behaviour at smaller magnetic fields, i.e. at larger $r_c$. This can be estimated quantitatively by calculating the cyclotron radius, given by
\begin{equation}
    r_c = \frac{v_F m^*}{e B} = \frac{\hbar \sqrt{n \pi}}{e B}
\end{equation}
where $v_F=10^6$ m s$^{-1}$ is the Fermi velocity in graphene, $m^*$ is the carrier effective mass,
$e$ is the charge of an electron, $B$ is the applied magnetic field, $\hbar$ is the reduced Planck's constant,
and $n$ is the carrier density.
For our devices, with a field of 1 T corresponding to the the field 
at which the transport features disappear and a carrier density of $10^{11}$ cm$^{-2}$ as 
determined by separate Hall measurements,
we find a cyclotron radius of 40 nm. This is in excellent agreement with the 50 nm radius
of the local delamination regions, as determined by SEM measurements.

           
           
Next we consider the gate voltage scale of the transport features.
Forward scattering off quasi-bound states in graphene quantum dots is also suppressed
when the energy of incident carriers matches the energy of a quasi-bound state\cite{silvestrov2007quantum}. 
We therefore expect the gate voltage scale of the transport features to match
the spacing between quasi-bound state energy levels.
Features in the transport data are spaced approximately 5V apart.
The change in the Fermi energy in graphene for a given change in gate voltage is given by
\begin{equation}
    \Delta E_F = e \cdot \alpha \cdot \Delta V_g
\end{equation}
where $\alpha = C_{bg} / C_Q$ is the capacitive lever arm of the back-gate, the capacitance of the backgate $C_{bg}$
is that of a SiO$_2$ parallel plate capacitor with an area of 200 $\mu$m$^2$ and a separation of 500 nm,
and the quantum capacitance of the graphene sheet is given by $C_Q = e^2 \cdot \rho(E_F)$ \cite{fang2007carrier}
(in the low temperature limit where $E_F \gg kT$). 
The density of states $\rho(E)$ is known for graphene\cite{CastroNeto2009}, and $E_F$ can be extracted from the
carrier density. Using a carrier density 
of $10^{11}$ cm$^{-2}$ as above this gives 
an energy scale for the observed transport features of approximately 40 meV.
Next we relate this experimental value to theoretical predictions.
            
For a parabolic potential of the form $V = − (x/x_0)^2 U/2$
theoretical results \cite{silvestrov2007quantum} give the following expression for the energy scale of the
quasi-bound states:
\begin{equation}
    E = \frac{\hbar v_F}{\xi} \;\;\; \text{with} \;\;\; \xi = \left[ \frac{\hbar v_F x_0^2}{U} \right]^{1/3}.
\end{equation}
As shown in Figure \ref{'fig:transport'}D, the potential profile in our devices is 
defined by the local delamination of graphene in the vicinity of the topographic features.
Approximating the potential profile in our devices as parabolic
we take $x_0=25$ nm to be half the radius of the
topographic features, and $U$ to be the change in potential created by the locally varying gate capacitance:
\begin{equation}
    U = e \cdot V_g \cdot (\alpha_\text{delaminated} - \alpha_\text{flat}).
\end{equation}
The capacitive lever arm for the flat region $\alpha_\text{flat}$ remains the same as above, and we model
the capacitance of the delaminated region as cylinder of air in series with the SiO$_2$ substrate.
The cylinder of air has radius $x_0$ and height 100 nm, where the latter figure
is half the height of the patterned features on the substrate. Taking $V_g = 22.5\text{V}$ 
as the midpoint of the region of interest we find a theoretically predicted energy scale of 57 meV for
the quasi-bound states in our device,
in good agreement with our experimental results. 

We note that the energy scale of the quasi-bound states depends on the height of the 
potential barrier as $E \propto U^{1/3} $. In our experimental configuration 
the height of the potential barrier depends 
in turn on the magnitude of the applied gate voltage as $ U \propto V_g$. 
The gate voltage in our measurements varies by a factor of two in the region of interest,
and so we expect the theoretically predicted energy scale to vary by a factor of $2^{1/3}$
over the course of the measurement, which precludes a more precise quantitative comparison.
It is also possible that local strain and pseudomagnetic fields affect transport in this regime. 
However, our data is consistent with dots having diameters commensurate with the delamination region, 
rather than localized around the pillars, which implies that delamination may dominate the physics we observe.

In summary, we describe a fabrication procedure for producing graphene devices on
substrates having an array of topographic features and we report experimental signatures
caused by these features. We find that the spacing of the topographic features
determines the adhesion behavior of the graphene; for large spacings partial
delamination creates strain in the graphene which we observe using optical measurements. 
Finally, we find that magneto-transport measurements
display features consistent with the presence of a locally varying potential, which we attribute to the 
variable gate capacitance induced by the local graphene delamination. 
These transport features are consistent with the presence of quasi-bound states in the graphene; the 
tunable nature of the graphene delamination offers an opportunity to deliberately engineer the properties
of these quasi-bound states.
With careful considerations of the topographic feature spacing and height
this technique can be adapted to length scales on the order of the coherence length in
graphene, thereby offering a novel method to explore correlated strain and electrostatic
potentials along with superlattice effects in graphene.

\section*{REFERENCES}
\bibliographystyle{aipnum4-1}
\bibliography{strainarray}

\begin{thebibliography}{40}%
\makeatletter
\providecommand \@ifxundefined [1]{%
 \@ifx{#1\undefined}
}%
\providecommand \@ifnum [1]{%
 \ifnum #1\expandafter \@firstoftwo
 \else \expandafter \@secondoftwo
 \fi
}%
\providecommand \@ifx [1]{%
 \ifx #1\expandafter \@firstoftwo
 \else \expandafter \@secondoftwo
 \fi
}%
\providecommand \natexlab [1]{#1}%
\providecommand \enquote  [1]{``#1''}%
\providecommand \bibnamefont  [1]{#1}%
\providecommand \bibfnamefont [1]{#1}%
\providecommand \citenamefont [1]{#1}%
\providecommand \href@noop [0]{\@secondoftwo}%
\providecommand \href [0]{\begingroup \@sanitize@url \@href}%
\providecommand \@href[1]{\@@startlink{#1}\@@href}%
\providecommand \@@href[1]{\endgroup#1\@@endlink}%
\providecommand \@sanitize@url [0]{\catcode `\\12\catcode `\$12\catcode
  `\&12\catcode `\#12\catcode `\^12\catcode `\_12\catcode `\%12\relax}%
\providecommand \@@startlink[1]{}%
\providecommand \@@endlink[0]{}%
\providecommand \url  [0]{\begingroup\@sanitize@url \@url }%
\providecommand \@url [1]{\endgroup\@href {#1}{\urlprefix }}%
\providecommand \urlprefix  [0]{URL }%
\providecommand \Eprint [0]{\href }%
\providecommand \doibase [0]{http://dx.doi.org/}%
\providecommand \selectlanguage [0]{\@gobble}%
\providecommand \bibinfo  [0]{\@secondoftwo}%
\providecommand \bibfield  [0]{\@secondoftwo}%
\providecommand \translation [1]{[#1]}%
\providecommand \BibitemOpen [0]{}%
\providecommand \bibitemStop [0]{}%
\providecommand \bibitemNoStop [0]{.\EOS\space}%
\providecommand \EOS [0]{\spacefactor3000\relax}%
\providecommand \BibitemShut  [1]{\csname bibitem#1\endcsname}%
\let\auto@bib@innerbib\@empty
\bibitem [{\citenamefont {Novoselov}\ \emph {et~al.}(2004)\citenamefont
  {Novoselov}, \citenamefont {Geim}, \citenamefont {Morozov}, \citenamefont
  {Jiang}, \citenamefont {Zhang}, \citenamefont {Dubonos}, \citenamefont
  {Grigorieva},\ and\ \citenamefont {Firsov}}]{novoselov2004electric}%
  \BibitemOpen
  \bibfield  {author} {\bibinfo {author} {\bibfnamefont {K.~S.}\ \bibnamefont
  {Novoselov}}, \bibinfo {author} {\bibfnamefont {A.~K.}\ \bibnamefont {Geim}},
  \bibinfo {author} {\bibfnamefont {S.~V.}\ \bibnamefont {Morozov}}, \bibinfo
  {author} {\bibfnamefont {D.}~\bibnamefont {Jiang}}, \bibinfo {author}
  {\bibfnamefont {Y.}~\bibnamefont {Zhang}}, \bibinfo {author} {\bibfnamefont
  {S.~V.}\ \bibnamefont {Dubonos}}, \bibinfo {author} {\bibfnamefont {I.~V.}\
  \bibnamefont {Grigorieva}}, \ and\ \bibinfo {author} {\bibfnamefont {A.~A.}\
  \bibnamefont {Firsov}},\ }\href@noop {} {\bibfield  {journal} {\bibinfo
  {journal} {Science}\ }\textbf {\bibinfo {volume} {306}},\ \bibinfo {pages}
  {666} (\bibinfo {year} {2004})}\BibitemShut {NoStop}%
\bibitem [{\citenamefont {Novoselov}\ \emph {et~al.}(2005)\citenamefont
  {Novoselov}, \citenamefont {Geim}, \citenamefont {Morozov}, \citenamefont
  {Jiang}, \citenamefont {Katsnelson}, \citenamefont {Grigorieva},
  \citenamefont {Dubonos},\ and\ \citenamefont {Firsov}}]{novoselov2005two}%
  \BibitemOpen
  \bibfield  {author} {\bibinfo {author} {\bibfnamefont {K.~S.}\ \bibnamefont
  {Novoselov}}, \bibinfo {author} {\bibfnamefont {A.~K.}\ \bibnamefont {Geim}},
  \bibinfo {author} {\bibfnamefont {S.~V.}\ \bibnamefont {Morozov}}, \bibinfo
  {author} {\bibfnamefont {D.}~\bibnamefont {Jiang}}, \bibinfo {author}
  {\bibfnamefont {M.}~\bibnamefont {Katsnelson}}, \bibinfo {author}
  {\bibfnamefont {I.~V.}\ \bibnamefont {Grigorieva}}, \bibinfo {author}
  {\bibfnamefont {S.}~\bibnamefont {Dubonos}}, \ and\ \bibinfo {author}
  {\bibfnamefont {A.}~\bibnamefont {Firsov}},\ }\href@noop {} {\bibfield
  {journal} {\bibinfo  {journal} {Nature}\ }\textbf {\bibinfo {volume} {438}},\
  \bibinfo {pages} {197} (\bibinfo {year} {2005})}\BibitemShut {NoStop}%
\bibitem [{\citenamefont {Zhang}\ \emph {et~al.}(2005)\citenamefont {Zhang},
  \citenamefont {Tan}, \citenamefont {Stormer},\ and\ \citenamefont
  {Kim}}]{zhang2005experimental}%
  \BibitemOpen
  \bibfield  {author} {\bibinfo {author} {\bibfnamefont {Y.}~\bibnamefont
  {Zhang}}, \bibinfo {author} {\bibfnamefont {Y.-W.}\ \bibnamefont {Tan}},
  \bibinfo {author} {\bibfnamefont {H.~L.}\ \bibnamefont {Stormer}}, \ and\
  \bibinfo {author} {\bibfnamefont {P.}~\bibnamefont {Kim}},\ }\href@noop {}
  {\bibfield  {journal} {\bibinfo  {journal} {Nature}\ }\textbf {\bibinfo
  {volume} {438}},\ \bibinfo {pages} {201} (\bibinfo {year}
  {2005})}\BibitemShut {NoStop}%
\bibitem [{\citenamefont {Geim}\ and\ \citenamefont
  {Novoselov}(2007)}]{geim2007rise}%
  \BibitemOpen
  \bibfield  {author} {\bibinfo {author} {\bibfnamefont {A.~K.}\ \bibnamefont
  {Geim}}\ and\ \bibinfo {author} {\bibfnamefont {K.~S.}\ \bibnamefont
  {Novoselov}},\ }\href@noop {} {\bibfield  {journal} {\bibinfo  {journal}
  {Nature Materials}\ }\textbf {\bibinfo {volume} {6}},\ \bibinfo {pages} {183}
  (\bibinfo {year} {2007})}\BibitemShut {NoStop}%
\bibitem [{\citenamefont {Guinea}, \citenamefont {Katsnelson},\ and\
  \citenamefont {Geim}(2010)}]{guinea2010energy}%
  \BibitemOpen
  \bibfield  {author} {\bibinfo {author} {\bibfnamefont {F.}~\bibnamefont
  {Guinea}}, \bibinfo {author} {\bibfnamefont {M.~I.}\ \bibnamefont
  {Katsnelson}}, \ and\ \bibinfo {author} {\bibfnamefont {A.~K.}\ \bibnamefont
  {Geim}},\ }\href@noop {} {\bibfield  {journal} {\bibinfo  {journal} {Nature
  Physics}\ }\textbf {\bibinfo {volume} {6}},\ \bibinfo {pages} {30} (\bibinfo
  {year} {2010})}\BibitemShut {NoStop}%
\bibitem [{\citenamefont {Zhou}\ \emph {et~al.}(2007)\citenamefont {Zhou},
  \citenamefont {Gweon}, \citenamefont {Fedorov}, \citenamefont {First},
  \citenamefont {De~Heer}, \citenamefont {Lee}, \citenamefont {Guinea},
  \citenamefont {{Castro Neto}},\ and\ \citenamefont
  {Lanzara}}]{zhou2007substrate}%
  \BibitemOpen
  \bibfield  {author} {\bibinfo {author} {\bibfnamefont {S.~Y.}\ \bibnamefont
  {Zhou}}, \bibinfo {author} {\bibfnamefont {G.-H.}\ \bibnamefont {Gweon}},
  \bibinfo {author} {\bibfnamefont {A.~V.}\ \bibnamefont {Fedorov}}, \bibinfo
  {author} {\bibfnamefont {P.~N.}\ \bibnamefont {First}}, \bibinfo {author}
  {\bibfnamefont {W.~A.}\ \bibnamefont {De~Heer}}, \bibinfo {author}
  {\bibfnamefont {D.-H.}\ \bibnamefont {Lee}}, \bibinfo {author} {\bibfnamefont
  {F.}~\bibnamefont {Guinea}}, \bibinfo {author} {\bibfnamefont {A.~H.}\
  \bibnamefont {{Castro Neto}}}, \ and\ \bibinfo {author} {\bibfnamefont
  {A.}~\bibnamefont {Lanzara}},\ }\href@noop {} {\bibfield  {journal} {\bibinfo
   {journal} {Nature Materials}\ }\textbf {\bibinfo {volume} {6}},\ \bibinfo
  {pages} {770} (\bibinfo {year} {2007})}\BibitemShut {NoStop}%
\bibitem [{\citenamefont {Tomori}\ \emph {et~al.}(2011)\citenamefont {Tomori},
  \citenamefont {Kanda}, \citenamefont {Goto}, \citenamefont {Ootuka},
  \citenamefont {Tsukagoshi}, \citenamefont {Moriyama}, \citenamefont
  {Watanabe},\ and\ \citenamefont {Tsuya}}]{Tomori2011}%
  \BibitemOpen
  \bibfield  {author} {\bibinfo {author} {\bibfnamefont {H.}~\bibnamefont
  {Tomori}}, \bibinfo {author} {\bibfnamefont {A.}~\bibnamefont {Kanda}},
  \bibinfo {author} {\bibfnamefont {H.}~\bibnamefont {Goto}}, \bibinfo {author}
  {\bibfnamefont {Y.}~\bibnamefont {Ootuka}}, \bibinfo {author} {\bibfnamefont
  {K.}~\bibnamefont {Tsukagoshi}}, \bibinfo {author} {\bibfnamefont
  {S.}~\bibnamefont {Moriyama}}, \bibinfo {author} {\bibfnamefont
  {E.}~\bibnamefont {Watanabe}}, \ and\ \bibinfo {author} {\bibfnamefont
  {D.}~\bibnamefont {Tsuya}},\ }\href
  {http://stacks.iop.org/1882-0786/4/i=7/a=075102} {\bibfield  {journal}
  {\bibinfo  {journal} {Applied Physics Express}\ }\textbf {\bibinfo {volume}
  {4}},\ \bibinfo {pages} {075102} (\bibinfo {year} {2011})}\BibitemShut
  {NoStop}%
\bibitem [{\citenamefont {Mi}\ \emph {et~al.}(2015)\citenamefont {Mi},
  \citenamefont {Mikael}, \citenamefont {Liu}, \citenamefont {Seo},
  \citenamefont {Gui}, \citenamefont {Ma}, \citenamefont {Nealey},\ and\
  \citenamefont {Ma}}]{mi2015creating}%
  \BibitemOpen
  \bibfield  {author} {\bibinfo {author} {\bibfnamefont {H.}~\bibnamefont
  {Mi}}, \bibinfo {author} {\bibfnamefont {S.}~\bibnamefont {Mikael}}, \bibinfo
  {author} {\bibfnamefont {C.-C.}\ \bibnamefont {Liu}}, \bibinfo {author}
  {\bibfnamefont {J.-H.}\ \bibnamefont {Seo}}, \bibinfo {author} {\bibfnamefont
  {G.}~\bibnamefont {Gui}}, \bibinfo {author} {\bibfnamefont {A.~L.}\
  \bibnamefont {Ma}}, \bibinfo {author} {\bibfnamefont {P.~F.}\ \bibnamefont
  {Nealey}}, \ and\ \bibinfo {author} {\bibfnamefont {Z.}~\bibnamefont {Ma}},\
  }\href@noop {} {\bibfield  {journal} {\bibinfo  {journal} {Applied Physics
  Letters}\ }\textbf {\bibinfo {volume} {107}},\ \bibinfo {pages} {143107}
  (\bibinfo {year} {2015})}\BibitemShut {NoStop}%
\bibitem [{\citenamefont {Babichev}\ \emph {et~al.}(2015)\citenamefont
  {Babichev}, \citenamefont {Rykov}, \citenamefont {Tchernycheva},
  \citenamefont {Smirnov}, \citenamefont {Davydov}, \citenamefont {Kumzerov},\
  and\ \citenamefont {Butko}}]{babichev2015influence}%
  \BibitemOpen
  \bibfield  {author} {\bibinfo {author} {\bibfnamefont {A.~V.}\ \bibnamefont
  {Babichev}}, \bibinfo {author} {\bibfnamefont {S.~A.}\ \bibnamefont {Rykov}},
  \bibinfo {author} {\bibfnamefont {M.}~\bibnamefont {Tchernycheva}}, \bibinfo
  {author} {\bibfnamefont {A.~N.}\ \bibnamefont {Smirnov}}, \bibinfo {author}
  {\bibfnamefont {V.~Y.}\ \bibnamefont {Davydov}}, \bibinfo {author}
  {\bibfnamefont {Y.~A.}\ \bibnamefont {Kumzerov}}, \ and\ \bibinfo {author}
  {\bibfnamefont {V.~Y.}\ \bibnamefont {Butko}},\ }\href@noop {} {\bibfield
  {journal} {\bibinfo  {journal} {ACS applied materials \& interfaces}\
  }\textbf {\bibinfo {volume} {8}},\ \bibinfo {pages} {240} (\bibinfo {year}
  {2015})}\BibitemShut {NoStop}%
\bibitem [{\citenamefont {Reserbat-Plantey}\ \emph {et~al.}(2014)\citenamefont
  {Reserbat-Plantey}, \citenamefont {Kalita}, \citenamefont {Han},
  \citenamefont {Ferlazzo}, \citenamefont {Autier-Laurent}, \citenamefont
  {Komatsu}, \citenamefont {Li}, \citenamefont {Weil}, \citenamefont {Ralko},
  \citenamefont {Marty} \emph {et~al.}}]{reserbat2014strain}%
  \BibitemOpen
  \bibfield  {author} {\bibinfo {author} {\bibfnamefont {A.}~\bibnamefont
  {Reserbat-Plantey}}, \bibinfo {author} {\bibfnamefont {D.}~\bibnamefont
  {Kalita}}, \bibinfo {author} {\bibfnamefont {Z.}~\bibnamefont {Han}},
  \bibinfo {author} {\bibfnamefont {L.}~\bibnamefont {Ferlazzo}}, \bibinfo
  {author} {\bibfnamefont {S.}~\bibnamefont {Autier-Laurent}}, \bibinfo
  {author} {\bibfnamefont {K.}~\bibnamefont {Komatsu}}, \bibinfo {author}
  {\bibfnamefont {C.}~\bibnamefont {Li}}, \bibinfo {author} {\bibfnamefont
  {R.}~\bibnamefont {Weil}}, \bibinfo {author} {\bibfnamefont {A.}~\bibnamefont
  {Ralko}}, \bibinfo {author} {\bibfnamefont {L.}~\bibnamefont {Marty}},  \emph
  {et~al.},\ }\href@noop {} {\bibfield  {journal} {\bibinfo  {journal} {Nano
  letters}\ }\textbf {\bibinfo {volume} {14}},\ \bibinfo {pages} {5044}
  (\bibinfo {year} {2014})}\BibitemShut {NoStop}%
\bibitem [{\citenamefont {Chiu}\ \emph {et~al.}(2010)\citenamefont {Chiu},
  \citenamefont {Perebeinos}, \citenamefont {Lin},\ and\ \citenamefont
  {Avouris}}]{chiu2010controllable}%
  \BibitemOpen
  \bibfield  {author} {\bibinfo {author} {\bibfnamefont {H.-Y.}\ \bibnamefont
  {Chiu}}, \bibinfo {author} {\bibfnamefont {V.}~\bibnamefont {Perebeinos}},
  \bibinfo {author} {\bibfnamefont {Y.-M.}\ \bibnamefont {Lin}}, \ and\
  \bibinfo {author} {\bibfnamefont {P.}~\bibnamefont {Avouris}},\ }\href@noop
  {} {\bibfield  {journal} {\bibinfo  {journal} {Nano Letters}\ }\textbf
  {\bibinfo {volume} {10}},\ \bibinfo {pages} {4634} (\bibinfo {year}
  {2010})}\BibitemShut {NoStop}%
\bibitem [{\citenamefont {Hinnefeld}\ \emph {et~al.}(2016)\citenamefont
  {Hinnefeld}, \citenamefont {Xu}, \citenamefont {Rogers}, \citenamefont
  {Pandya}, \citenamefont {Shim}, \citenamefont {Martin},\ and\ \citenamefont
  {Mason}}]{hinnefeld2016single}%
  \BibitemOpen
  \bibfield  {author} {\bibinfo {author} {\bibfnamefont {J.~H.}\ \bibnamefont
  {Hinnefeld}}, \bibinfo {author} {\bibfnamefont {R.}~\bibnamefont {Xu}},
  \bibinfo {author} {\bibfnamefont {S.}~\bibnamefont {Rogers}}, \bibinfo
  {author} {\bibfnamefont {S.}~\bibnamefont {Pandya}}, \bibinfo {author}
  {\bibfnamefont {M.}~\bibnamefont {Shim}}, \bibinfo {author} {\bibfnamefont
  {L.~W.}\ \bibnamefont {Martin}}, \ and\ \bibinfo {author} {\bibfnamefont
  {N.}~\bibnamefont {Mason}},\ }\href@noop {} {\bibfield  {journal} {\bibinfo
  {journal} {Applied Physics Letters}\ }\textbf {\bibinfo {volume} {108}},\
  \bibinfo {pages} {203109} (\bibinfo {year} {2016})}\BibitemShut {NoStop}%
\bibitem [{\citenamefont {Pereira}\ and\ \citenamefont {{Castro
  Neto}}(2009)}]{pereira2009strain}%
  \BibitemOpen
  \bibfield  {author} {\bibinfo {author} {\bibfnamefont {V.~M.}\ \bibnamefont
  {Pereira}}\ and\ \bibinfo {author} {\bibfnamefont {A.~H.}\ \bibnamefont
  {{Castro Neto}}},\ }\href@noop {} {\bibfield  {journal} {\bibinfo  {journal}
  {Physical Review Letters}\ }\textbf {\bibinfo {volume} {103}},\ \bibinfo
  {pages} {046801} (\bibinfo {year} {2009})}\BibitemShut {NoStop}%
\bibitem [{\citenamefont {Ni}\ \emph {et~al.}(2008)\citenamefont {Ni},
  \citenamefont {Yu}, \citenamefont {Lu}, \citenamefont {Wang}, \citenamefont
  {Feng},\ and\ \citenamefont {Shen}}]{ni2008uniaxial}%
  \BibitemOpen
  \bibfield  {author} {\bibinfo {author} {\bibfnamefont {Z.~H.}\ \bibnamefont
  {Ni}}, \bibinfo {author} {\bibfnamefont {T.}~\bibnamefont {Yu}}, \bibinfo
  {author} {\bibfnamefont {Y.~H.}\ \bibnamefont {Lu}}, \bibinfo {author}
  {\bibfnamefont {Y.~Y.}\ \bibnamefont {Wang}}, \bibinfo {author}
  {\bibfnamefont {Y.~P.}\ \bibnamefont {Feng}}, \ and\ \bibinfo {author}
  {\bibfnamefont {Z.~X.}\ \bibnamefont {Shen}},\ }\href@noop {} {\bibfield
  {journal} {\bibinfo  {journal} {ACS Nano}\ }\textbf {\bibinfo {volume} {2}},\
  \bibinfo {pages} {2301} (\bibinfo {year} {2008})}\BibitemShut {NoStop}%
\bibitem [{\citenamefont {Levy}\ \emph {et~al.}(2010)\citenamefont {Levy},
  \citenamefont {Burke}, \citenamefont {Meaker}, \citenamefont {Panlasigui},
  \citenamefont {Zettl}, \citenamefont {Guinea}, \citenamefont {{Castro
  Neto}},\ and\ \citenamefont {Crommie}}]{levy2010strain}%
  \BibitemOpen
  \bibfield  {author} {\bibinfo {author} {\bibfnamefont {N.}~\bibnamefont
  {Levy}}, \bibinfo {author} {\bibfnamefont {S.~A.}\ \bibnamefont {Burke}},
  \bibinfo {author} {\bibfnamefont {K.~L.}\ \bibnamefont {Meaker}}, \bibinfo
  {author} {\bibfnamefont {M.}~\bibnamefont {Panlasigui}}, \bibinfo {author}
  {\bibfnamefont {A.}~\bibnamefont {Zettl}}, \bibinfo {author} {\bibfnamefont
  {F.}~\bibnamefont {Guinea}}, \bibinfo {author} {\bibfnamefont {A.~H.}\
  \bibnamefont {{Castro Neto}}}, \ and\ \bibinfo {author} {\bibfnamefont
  {M.~F.}\ \bibnamefont {Crommie}},\ }\href@noop {} {\bibfield  {journal}
  {\bibinfo  {journal} {Science}\ }\textbf {\bibinfo {volume} {329}},\ \bibinfo
  {pages} {544} (\bibinfo {year} {2010})}\BibitemShut {NoStop}%
\bibitem [{\citenamefont {Mohiuddin}\ \emph {et~al.}(2009)\citenamefont
  {Mohiuddin}, \citenamefont {Lombardo}, \citenamefont {Nair}, \citenamefont
  {Bonetti}, \citenamefont {Savini}, \citenamefont {Jalil}, \citenamefont
  {Bonini}, \citenamefont {Basko}, \citenamefont {Galiotis}, \citenamefont
  {Marzari} \emph {et~al.}}]{mohiuddin2009uniaxial}%
  \BibitemOpen
  \bibfield  {author} {\bibinfo {author} {\bibfnamefont {T.~M.~G.}\
  \bibnamefont {Mohiuddin}}, \bibinfo {author} {\bibfnamefont {A.}~\bibnamefont
  {Lombardo}}, \bibinfo {author} {\bibfnamefont {R.~R.}\ \bibnamefont {Nair}},
  \bibinfo {author} {\bibfnamefont {A.}~\bibnamefont {Bonetti}}, \bibinfo
  {author} {\bibfnamefont {G.}~\bibnamefont {Savini}}, \bibinfo {author}
  {\bibfnamefont {R.}~\bibnamefont {Jalil}}, \bibinfo {author} {\bibfnamefont
  {N.}~\bibnamefont {Bonini}}, \bibinfo {author} {\bibfnamefont {D.~M.}\
  \bibnamefont {Basko}}, \bibinfo {author} {\bibfnamefont {C.}~\bibnamefont
  {Galiotis}}, \bibinfo {author} {\bibfnamefont {N.}~\bibnamefont {Marzari}},
  \emph {et~al.},\ }\href@noop {} {\bibfield  {journal} {\bibinfo  {journal}
  {Physical Review B}\ }\textbf {\bibinfo {volume} {79}},\ \bibinfo {pages}
  {205433} (\bibinfo {year} {2009})}\BibitemShut {NoStop}%
\bibitem [{\citenamefont {Gill}\ \emph {et~al.}(2015)\citenamefont {Gill},
  \citenamefont {Hinnefeld}, \citenamefont {Zhu}, \citenamefont {Swanson},
  \citenamefont {Li},\ and\ \citenamefont {Mason}}]{gill2015mechanical}%
  \BibitemOpen
  \bibfield  {author} {\bibinfo {author} {\bibfnamefont {S.~T.}\ \bibnamefont
  {Gill}}, \bibinfo {author} {\bibfnamefont {J.~H.}\ \bibnamefont {Hinnefeld}},
  \bibinfo {author} {\bibfnamefont {S.}~\bibnamefont {Zhu}}, \bibinfo {author}
  {\bibfnamefont {W.~J.}\ \bibnamefont {Swanson}}, \bibinfo {author}
  {\bibfnamefont {T.}~\bibnamefont {Li}}, \ and\ \bibinfo {author}
  {\bibfnamefont {N.}~\bibnamefont {Mason}},\ }\href@noop {} {\bibfield
  {journal} {\bibinfo  {journal} {ACS Nano}\ }\textbf {\bibinfo {volume} {9}},\
  \bibinfo {pages} {5799} (\bibinfo {year} {2015})}\BibitemShut {NoStop}%
\bibitem [{\citenamefont {Lee}\ \emph {et~al.}(2012)\citenamefont {Lee},
  \citenamefont {Ahn}, \citenamefont {Shim}, \citenamefont {Lee},\ and\
  \citenamefont {Ryu}}]{lee2012optical}%
  \BibitemOpen
  \bibfield  {author} {\bibinfo {author} {\bibfnamefont {J.~E.}\ \bibnamefont
  {Lee}}, \bibinfo {author} {\bibfnamefont {G.}~\bibnamefont {Ahn}}, \bibinfo
  {author} {\bibfnamefont {J.}~\bibnamefont {Shim}}, \bibinfo {author}
  {\bibfnamefont {Y.~S.}\ \bibnamefont {Lee}}, \ and\ \bibinfo {author}
  {\bibfnamefont {S.}~\bibnamefont {Ryu}},\ }\href@noop {} {\bibfield
  {journal} {\bibinfo  {journal} {Nature Communications}\ }\textbf {\bibinfo
  {volume} {3}},\ \bibinfo {pages} {1024} (\bibinfo {year} {2012})}\BibitemShut
  {NoStop}%
\bibitem [{\citenamefont {Li}\ \emph {et~al.}(2009{\natexlab{a}})\citenamefont
  {Li}, \citenamefont {Cai}, \citenamefont {An}, \citenamefont {Kim},
  \citenamefont {Nah}, \citenamefont {Yang}, \citenamefont {Piner},
  \citenamefont {Velamakanni}, \citenamefont {Jung}, \citenamefont {Tutuc},
  \citenamefont {Banerjee}, \citenamefont {Colombo},\ and\ \citenamefont
  {Ruoff}}]{Li2009}%
  \BibitemOpen
  \bibfield  {author} {\bibinfo {author} {\bibfnamefont {X.}~\bibnamefont
  {Li}}, \bibinfo {author} {\bibfnamefont {W.}~\bibnamefont {Cai}}, \bibinfo
  {author} {\bibfnamefont {J.}~\bibnamefont {An}}, \bibinfo {author}
  {\bibfnamefont {S.}~\bibnamefont {Kim}}, \bibinfo {author} {\bibfnamefont
  {J.}~\bibnamefont {Nah}}, \bibinfo {author} {\bibfnamefont {D.}~\bibnamefont
  {Yang}}, \bibinfo {author} {\bibfnamefont {R.}~\bibnamefont {Piner}},
  \bibinfo {author} {\bibfnamefont {A.}~\bibnamefont {Velamakanni}}, \bibinfo
  {author} {\bibfnamefont {I.}~\bibnamefont {Jung}}, \bibinfo {author}
  {\bibfnamefont {E.}~\bibnamefont {Tutuc}}, \bibinfo {author} {\bibfnamefont
  {S.~K.}\ \bibnamefont {Banerjee}}, \bibinfo {author} {\bibfnamefont
  {L.}~\bibnamefont {Colombo}}, \ and\ \bibinfo {author} {\bibfnamefont
  {R.~S.}\ \bibnamefont {Ruoff}},\ }\href {\doibase 10.1126/science.1171245}
  {\bibfield  {journal} {\bibinfo  {journal} {Science}\ }\textbf {\bibinfo
  {volume} {324}},\ \bibinfo {pages} {1312} (\bibinfo {year}
  {2009}{\natexlab{a}})}\BibitemShut {NoStop}%
\bibitem [{\citenamefont {Li}\ \emph {et~al.}(2009{\natexlab{b}})\citenamefont
  {Li}, \citenamefont {Zhu}, \citenamefont {Cai}, \citenamefont {Borysiak},
  \citenamefont {Han}, \citenamefont {Chen}, \citenamefont {Piner},
  \citenamefont {Colombo},\ and\ \citenamefont {Ruoff}}]{li2009transfer}%
  \BibitemOpen
  \bibfield  {author} {\bibinfo {author} {\bibfnamefont {X.}~\bibnamefont
  {Li}}, \bibinfo {author} {\bibfnamefont {Y.}~\bibnamefont {Zhu}}, \bibinfo
  {author} {\bibfnamefont {W.}~\bibnamefont {Cai}}, \bibinfo {author}
  {\bibfnamefont {M.}~\bibnamefont {Borysiak}}, \bibinfo {author}
  {\bibfnamefont {B.}~\bibnamefont {Han}}, \bibinfo {author} {\bibfnamefont
  {D.}~\bibnamefont {Chen}}, \bibinfo {author} {\bibfnamefont {R.~D.}\
  \bibnamefont {Piner}}, \bibinfo {author} {\bibfnamefont {L.}~\bibnamefont
  {Colombo}}, \ and\ \bibinfo {author} {\bibfnamefont {R.~S.}\ \bibnamefont
  {Ruoff}},\ }\href@noop {} {\bibfield  {journal} {\bibinfo  {journal} {Nano
  Letters}\ }\textbf {\bibinfo {volume} {9}},\ \bibinfo {pages} {4359}
  (\bibinfo {year} {2009}{\natexlab{b}})}\BibitemShut {NoStop}%
\bibitem [{\citenamefont {Yamamoto}\ \emph {et~al.}(2012)\citenamefont
  {Yamamoto}, \citenamefont {Pierre-Louis}, \citenamefont {Huang},
  \citenamefont {Fuhrer}, \citenamefont {Einstein},\ and\ \citenamefont
  {Cullen}}]{yamamoto2012princess}%
  \BibitemOpen
  \bibfield  {author} {\bibinfo {author} {\bibfnamefont {M.}~\bibnamefont
  {Yamamoto}}, \bibinfo {author} {\bibfnamefont {O.}~\bibnamefont
  {Pierre-Louis}}, \bibinfo {author} {\bibfnamefont {J.}~\bibnamefont {Huang}},
  \bibinfo {author} {\bibfnamefont {M.~S.}\ \bibnamefont {Fuhrer}}, \bibinfo
  {author} {\bibfnamefont {T.~L.}\ \bibnamefont {Einstein}}, \ and\ \bibinfo
  {author} {\bibfnamefont {W.~G.}\ \bibnamefont {Cullen}},\ }\href@noop {}
  {\bibfield  {journal} {\bibinfo  {journal} {Physical Review X}\ }\textbf
  {\bibinfo {volume} {2}},\ \bibinfo {pages} {041018} (\bibinfo {year}
  {2012})}\BibitemShut {NoStop}%
\bibitem [{\citenamefont {Ferrari}\ \emph {et~al.}(2006)\citenamefont
  {Ferrari}, \citenamefont {Meyer}, \citenamefont {Scardaci}, \citenamefont
  {Casiraghi}, \citenamefont {Lazzeri}, \citenamefont {Mauri}, \citenamefont
  {Piscanec}, \citenamefont {Jiang}, \citenamefont {Novoselov}, \citenamefont
  {Roth} \emph {et~al.}}]{ferrari2006raman}%
  \BibitemOpen
  \bibfield  {author} {\bibinfo {author} {\bibfnamefont {A.}~\bibnamefont
  {Ferrari}}, \bibinfo {author} {\bibfnamefont {J.}~\bibnamefont {Meyer}},
  \bibinfo {author} {\bibfnamefont {V.}~\bibnamefont {Scardaci}}, \bibinfo
  {author} {\bibfnamefont {C.}~\bibnamefont {Casiraghi}}, \bibinfo {author}
  {\bibfnamefont {M.}~\bibnamefont {Lazzeri}}, \bibinfo {author} {\bibfnamefont
  {F.}~\bibnamefont {Mauri}}, \bibinfo {author} {\bibfnamefont
  {S.}~\bibnamefont {Piscanec}}, \bibinfo {author} {\bibfnamefont
  {D.}~\bibnamefont {Jiang}}, \bibinfo {author} {\bibfnamefont
  {K.}~\bibnamefont {Novoselov}}, \bibinfo {author} {\bibfnamefont
  {S.}~\bibnamefont {Roth}},  \emph {et~al.},\ }\href@noop {} {\bibfield
  {journal} {\bibinfo  {journal} {Physical Review Letters}\ }\textbf {\bibinfo
  {volume} {97}},\ \bibinfo {pages} {187401} (\bibinfo {year}
  {2006})}\BibitemShut {NoStop}%
\bibitem [{\citenamefont {Scharfenberg}\ \emph {et~al.}(2012)\citenamefont
  {Scharfenberg}, \citenamefont {Mansukhani}, \citenamefont {Chialvo},
  \citenamefont {Weaver},\ and\ \citenamefont
  {Mason}}]{scharfenberg2012observation}%
  \BibitemOpen
  \bibfield  {author} {\bibinfo {author} {\bibfnamefont {S.}~\bibnamefont
  {Scharfenberg}}, \bibinfo {author} {\bibfnamefont {N.}~\bibnamefont
  {Mansukhani}}, \bibinfo {author} {\bibfnamefont {C.}~\bibnamefont {Chialvo}},
  \bibinfo {author} {\bibfnamefont {R.~L.}\ \bibnamefont {Weaver}}, \ and\
  \bibinfo {author} {\bibfnamefont {N.}~\bibnamefont {Mason}},\ }\href@noop {}
  {\bibfield  {journal} {\bibinfo  {journal} {Applied Physics Letters}\
  }\textbf {\bibinfo {volume} {100}},\ \bibinfo {pages} {021910} (\bibinfo
  {year} {2012})}\BibitemShut {NoStop}%
\bibitem [{\citenamefont {Reina}\ \emph {et~al.}(2008)\citenamefont {Reina},
  \citenamefont {Jia}, \citenamefont {Ho}, \citenamefont {Nezich},
  \citenamefont {Son}, \citenamefont {Bulovic}, \citenamefont {Dresselhaus},\
  and\ \citenamefont {Kong}}]{reina2008large}%
  \BibitemOpen
  \bibfield  {author} {\bibinfo {author} {\bibfnamefont {A.}~\bibnamefont
  {Reina}}, \bibinfo {author} {\bibfnamefont {X.}~\bibnamefont {Jia}}, \bibinfo
  {author} {\bibfnamefont {J.}~\bibnamefont {Ho}}, \bibinfo {author}
  {\bibfnamefont {D.}~\bibnamefont {Nezich}}, \bibinfo {author} {\bibfnamefont
  {H.}~\bibnamefont {Son}}, \bibinfo {author} {\bibfnamefont {V.}~\bibnamefont
  {Bulovic}}, \bibinfo {author} {\bibfnamefont {M.~S.}\ \bibnamefont
  {Dresselhaus}}, \ and\ \bibinfo {author} {\bibfnamefont {J.}~\bibnamefont
  {Kong}},\ }\href@noop {} {\bibfield  {journal} {\bibinfo  {journal} {Nano
  Letters}\ }\textbf {\bibinfo {volume} {9}},\ \bibinfo {pages} {30} (\bibinfo
  {year} {2008})}\BibitemShut {NoStop}%
\bibitem [{\citenamefont {Casiraghi}\ \emph {et~al.}(2007)\citenamefont
  {Casiraghi}, \citenamefont {Pisana}, \citenamefont {Novoselov}, \citenamefont
  {Geim},\ and\ \citenamefont {Ferrari}}]{casiraghi2007raman}%
  \BibitemOpen
  \bibfield  {author} {\bibinfo {author} {\bibfnamefont {C.}~\bibnamefont
  {Casiraghi}}, \bibinfo {author} {\bibfnamefont {S.}~\bibnamefont {Pisana}},
  \bibinfo {author} {\bibfnamefont {K.~S.}\ \bibnamefont {Novoselov}}, \bibinfo
  {author} {\bibfnamefont {A.~K.}\ \bibnamefont {Geim}}, \ and\ \bibinfo
  {author} {\bibfnamefont {A.~C.}\ \bibnamefont {Ferrari}},\ }\href@noop {}
  {\bibfield  {journal} {\bibinfo  {journal} {Applied Physics Letters}\
  }\textbf {\bibinfo {volume} {91}},\ \bibinfo {pages} {233108} (\bibinfo
  {year} {2007})}\BibitemShut {NoStop}%
\bibitem [{\citenamefont {Zabel}\ \emph {et~al.}(2012)\citenamefont {Zabel},
  \citenamefont {Nair}, \citenamefont {Ott}, \citenamefont {Georgiou},
  \citenamefont {Geim}, \citenamefont {Novoselov},\ and\ \citenamefont
  {Casiraghi}}]{zabel2012raman}%
  \BibitemOpen
  \bibfield  {author} {\bibinfo {author} {\bibfnamefont {J.}~\bibnamefont
  {Zabel}}, \bibinfo {author} {\bibfnamefont {R.~R.}\ \bibnamefont {Nair}},
  \bibinfo {author} {\bibfnamefont {A.}~\bibnamefont {Ott}}, \bibinfo {author}
  {\bibfnamefont {T.}~\bibnamefont {Georgiou}}, \bibinfo {author}
  {\bibfnamefont {A.~K.}\ \bibnamefont {Geim}}, \bibinfo {author}
  {\bibfnamefont {K.~S.}\ \bibnamefont {Novoselov}}, \ and\ \bibinfo {author}
  {\bibfnamefont {C.}~\bibnamefont {Casiraghi}},\ }\href@noop {} {\bibfield
  {journal} {\bibinfo  {journal} {Nano Letters}\ }\textbf {\bibinfo {volume}
  {12}},\ \bibinfo {pages} {617} (\bibinfo {year} {2012})}\BibitemShut
  {NoStop}%
\bibitem [{\citenamefont {Metzger}\ \emph {et~al.}(2009)\citenamefont
  {Metzger}, \citenamefont {R{\'e}mi}, \citenamefont {Liu}, \citenamefont
  {Kusminskiy}, \citenamefont {{Castro Neto}}, \citenamefont {Swan},\ and\
  \citenamefont {Goldberg}}]{metzger2009biaxial}%
  \BibitemOpen
  \bibfield  {author} {\bibinfo {author} {\bibfnamefont {C.}~\bibnamefont
  {Metzger}}, \bibinfo {author} {\bibfnamefont {S.}~\bibnamefont {R{\'e}mi}},
  \bibinfo {author} {\bibfnamefont {M.}~\bibnamefont {Liu}}, \bibinfo {author}
  {\bibfnamefont {S.~V.}\ \bibnamefont {Kusminskiy}}, \bibinfo {author}
  {\bibfnamefont {A.~H.}\ \bibnamefont {{Castro Neto}}}, \bibinfo {author}
  {\bibfnamefont {A.~K.}\ \bibnamefont {Swan}}, \ and\ \bibinfo {author}
  {\bibfnamefont {B.~B.}\ \bibnamefont {Goldberg}},\ }\href@noop {} {\bibfield
  {journal} {\bibinfo  {journal} {Nano Letters}\ }\textbf {\bibinfo {volume}
  {10}},\ \bibinfo {pages} {6} (\bibinfo {year} {2009})}\BibitemShut {NoStop}%
\bibitem [{\citenamefont {Ding}\ \emph {et~al.}(2010)\citenamefont {Ding},
  \citenamefont {Ji}, \citenamefont {Chen}, \citenamefont {Herklotz},
  \citenamefont {Dörr}, \citenamefont {Mei}, \citenamefont {Rastelli},\ and\
  \citenamefont {Schmidt}}]{ding2010stretchable}%
  \BibitemOpen
  \bibfield  {author} {\bibinfo {author} {\bibfnamefont {F.}~\bibnamefont
  {Ding}}, \bibinfo {author} {\bibfnamefont {H.}~\bibnamefont {Ji}}, \bibinfo
  {author} {\bibfnamefont {Y.}~\bibnamefont {Chen}}, \bibinfo {author}
  {\bibfnamefont {A.}~\bibnamefont {Herklotz}}, \bibinfo {author}
  {\bibfnamefont {K.}~\bibnamefont {Dörr}}, \bibinfo {author} {\bibfnamefont
  {Y.}~\bibnamefont {Mei}}, \bibinfo {author} {\bibfnamefont {A.}~\bibnamefont
  {Rastelli}}, \ and\ \bibinfo {author} {\bibfnamefont {O.~G.}\ \bibnamefont
  {Schmidt}},\ }\href@noop {} {\bibfield  {journal} {\bibinfo  {journal} {Nano
  Letters}\ }\textbf {\bibinfo {volume} {10}},\ \bibinfo {pages} {3453}
  (\bibinfo {year} {2010})}\BibitemShut {NoStop}%
\bibitem [{\citenamefont {Mohr}, \citenamefont {Maultzsch},\ and\ \citenamefont
  {Thomsen}(2010)}]{mohr2010splitting}%
  \BibitemOpen
  \bibfield  {author} {\bibinfo {author} {\bibfnamefont {M.}~\bibnamefont
  {Mohr}}, \bibinfo {author} {\bibfnamefont {J.}~\bibnamefont {Maultzsch}}, \
  and\ \bibinfo {author} {\bibfnamefont {C.}~\bibnamefont {Thomsen}},\
  }\href@noop {} {\bibfield  {journal} {\bibinfo  {journal} {Physical Review
  B}\ }\textbf {\bibinfo {volume} {82}},\ \bibinfo {pages} {201409} (\bibinfo
  {year} {2010})}\BibitemShut {NoStop}%
\bibitem [{\citenamefont {Katsnelson}, \citenamefont {Novoselov},\ and\
  \citenamefont {Geim}(2006)}]{katsnelson2006chiral}%
  \BibitemOpen
  \bibfield  {author} {\bibinfo {author} {\bibfnamefont {M.}~\bibnamefont
  {Katsnelson}}, \bibinfo {author} {\bibfnamefont {K.~S.}\ \bibnamefont
  {Novoselov}}, \ and\ \bibinfo {author} {\bibfnamefont {A.~K.}\ \bibnamefont
  {Geim}},\ }\href@noop {} {\bibfield  {journal} {\bibinfo  {journal} {Nature
  Physics}\ }\textbf {\bibinfo {volume} {2}},\ \bibinfo {pages} {620} (\bibinfo
  {year} {2006})}\BibitemShut {NoStop}%
\bibitem [{\citenamefont {Young}\ and\ \citenamefont
  {Kim}(2009)}]{young2009quantum}%
  \BibitemOpen
  \bibfield  {author} {\bibinfo {author} {\bibfnamefont {A.~F.}\ \bibnamefont
  {Young}}\ and\ \bibinfo {author} {\bibfnamefont {P.}~\bibnamefont {Kim}},\
  }\href@noop {} {\bibfield  {journal} {\bibinfo  {journal} {Nature Physics}\
  }\textbf {\bibinfo {volume} {5}},\ \bibinfo {pages} {222} (\bibinfo {year}
  {2009})}\BibitemShut {NoStop}%
\bibitem [{\citenamefont {Silvestrov}\ and\ \citenamefont
  {Efetov}(2007)}]{silvestrov2007quantum}%
  \BibitemOpen
  \bibfield  {author} {\bibinfo {author} {\bibfnamefont {P.~G.}\ \bibnamefont
  {Silvestrov}}\ and\ \bibinfo {author} {\bibfnamefont {K.~B.}\ \bibnamefont
  {Efetov}},\ }\href@noop {} {\bibfield  {journal} {\bibinfo  {journal}
  {Physical Review Letters}\ }\textbf {\bibinfo {volume} {98}},\ \bibinfo
  {pages} {016802} (\bibinfo {year} {2007})}\BibitemShut {NoStop}%
\bibitem [{\citenamefont {Matulis}\ and\ \citenamefont
  {Peeters}(2008)}]{matulis2008quasibound}%
  \BibitemOpen
  \bibfield  {author} {\bibinfo {author} {\bibfnamefont {A.}~\bibnamefont
  {Matulis}}\ and\ \bibinfo {author} {\bibfnamefont {F.~M.}\ \bibnamefont
  {Peeters}},\ }\href@noop {} {\bibfield  {journal} {\bibinfo  {journal}
  {Physical Review B}\ }\textbf {\bibinfo {volume} {77}},\ \bibinfo {pages}
  {115423} (\bibinfo {year} {2008})}\BibitemShut {NoStop}%
\bibitem [{\citenamefont {Hewageegana}\ and\ \citenamefont
  {Apalkov}(2008)}]{hewageegana2008electron}%
  \BibitemOpen
  \bibfield  {author} {\bibinfo {author} {\bibfnamefont {P.}~\bibnamefont
  {Hewageegana}}\ and\ \bibinfo {author} {\bibfnamefont {V.}~\bibnamefont
  {Apalkov}},\ }\href@noop {} {\bibfield  {journal} {\bibinfo  {journal}
  {Physical Review B}\ }\textbf {\bibinfo {volume} {77}},\ \bibinfo {pages}
  {245426} (\bibinfo {year} {2008})}\BibitemShut {NoStop}%
\bibitem [{\citenamefont {Chen}, \citenamefont {Apalkov},\ and\ \citenamefont
  {Chakraborty}(2007)}]{chen2007fock}%
  \BibitemOpen
  \bibfield  {author} {\bibinfo {author} {\bibfnamefont {H.-Y.}\ \bibnamefont
  {Chen}}, \bibinfo {author} {\bibfnamefont {V.}~\bibnamefont {Apalkov}}, \
  and\ \bibinfo {author} {\bibfnamefont {T.}~\bibnamefont {Chakraborty}},\
  }\href@noop {} {\bibfield  {journal} {\bibinfo  {journal} {Physical Review
  Letters}\ }\textbf {\bibinfo {volume} {98}},\ \bibinfo {pages} {186803}
  (\bibinfo {year} {2007})}\BibitemShut {NoStop}%
\bibitem [{\citenamefont {Heinisch}, \citenamefont {Bronold},\ and\
  \citenamefont {Fehske}(2013)}]{heinisch2013mie}%
  \BibitemOpen
  \bibfield  {author} {\bibinfo {author} {\bibfnamefont {R.~L.}\ \bibnamefont
  {Heinisch}}, \bibinfo {author} {\bibfnamefont {F.~X.}\ \bibnamefont
  {Bronold}}, \ and\ \bibinfo {author} {\bibfnamefont {H.}~\bibnamefont
  {Fehske}},\ }\href@noop {} {\bibfield  {journal} {\bibinfo  {journal}
  {Physical Review B}\ }\textbf {\bibinfo {volume} {87}},\ \bibinfo {pages}
  {155409} (\bibinfo {year} {2013})}\BibitemShut {NoStop}%
\bibitem [{\citenamefont {Berger}\ \emph {et~al.}(2006)\citenamefont {Berger},
  \citenamefont {Song}, \citenamefont {Li}, \citenamefont {Wu}, \citenamefont
  {Brown}, \citenamefont {Naud}, \citenamefont {Mayou}, \citenamefont {Li},
  \citenamefont {Hass}, \citenamefont {Marchenkov} \emph
  {et~al.}}]{berger2006electronic}%
  \BibitemOpen
  \bibfield  {author} {\bibinfo {author} {\bibfnamefont {C.}~\bibnamefont
  {Berger}}, \bibinfo {author} {\bibfnamefont {Z.}~\bibnamefont {Song}},
  \bibinfo {author} {\bibfnamefont {X.}~\bibnamefont {Li}}, \bibinfo {author}
  {\bibfnamefont {X.}~\bibnamefont {Wu}}, \bibinfo {author} {\bibfnamefont
  {N.}~\bibnamefont {Brown}}, \bibinfo {author} {\bibfnamefont
  {C.}~\bibnamefont {Naud}}, \bibinfo {author} {\bibfnamefont {D.}~\bibnamefont
  {Mayou}}, \bibinfo {author} {\bibfnamefont {T.}~\bibnamefont {Li}}, \bibinfo
  {author} {\bibfnamefont {J.}~\bibnamefont {Hass}}, \bibinfo {author}
  {\bibfnamefont {A.~N.}\ \bibnamefont {Marchenkov}},  \emph {et~al.},\
  }\href@noop {} {\bibfield  {journal} {\bibinfo  {journal} {Science}\ }\textbf
  {\bibinfo {volume} {312}},\ \bibinfo {pages} {1191} (\bibinfo {year}
  {2006})}\BibitemShut {NoStop}%
\bibitem [{\citenamefont {Bolotin}\ \emph {et~al.}(2009)\citenamefont
  {Bolotin}, \citenamefont {Ghahari}, \citenamefont {Shulman}, \citenamefont
  {Stormer},\ and\ \citenamefont {Kim}}]{bolotin2009observation}%
  \BibitemOpen
  \bibfield  {author} {\bibinfo {author} {\bibfnamefont {K.~I.}\ \bibnamefont
  {Bolotin}}, \bibinfo {author} {\bibfnamefont {F.}~\bibnamefont {Ghahari}},
  \bibinfo {author} {\bibfnamefont {M.~D.}\ \bibnamefont {Shulman}}, \bibinfo
  {author} {\bibfnamefont {H.~L.}\ \bibnamefont {Stormer}}, \ and\ \bibinfo
  {author} {\bibfnamefont {P.}~\bibnamefont {Kim}},\ }\href@noop {} {\bibfield
  {journal} {\bibinfo  {journal} {Nature}\ }\textbf {\bibinfo {volume} {462}},\
  \bibinfo {pages} {196} (\bibinfo {year} {2009})}\BibitemShut {NoStop}%
\bibitem [{\citenamefont {Fang}\ \emph {et~al.}(2007)\citenamefont {Fang},
  \citenamefont {Konar}, \citenamefont {Xing},\ and\ \citenamefont
  {Jena}}]{fang2007carrier}%
  \BibitemOpen
  \bibfield  {author} {\bibinfo {author} {\bibfnamefont {T.}~\bibnamefont
  {Fang}}, \bibinfo {author} {\bibfnamefont {A.}~\bibnamefont {Konar}},
  \bibinfo {author} {\bibfnamefont {H.}~\bibnamefont {Xing}}, \ and\ \bibinfo
  {author} {\bibfnamefont {D.}~\bibnamefont {Jena}},\ }\href@noop {} {\bibfield
   {journal} {\bibinfo  {journal} {Applied Physics Letters}\ }\textbf {\bibinfo
  {volume} {91}},\ \bibinfo {pages} {092109} (\bibinfo {year}
  {2007})}\BibitemShut {NoStop}%
\bibitem [{\citenamefont {{Castro Neto}}\ \emph {et~al.}(2009)\citenamefont
  {{Castro Neto}}, \citenamefont {Peres}, \citenamefont {Novoselov},\ and\
  \citenamefont {Geim}}]{CastroNeto2009}%
  \BibitemOpen
  \bibfield  {author} {\bibinfo {author} {\bibfnamefont {A.~H.}\ \bibnamefont
  {{Castro Neto}}}, \bibinfo {author} {\bibfnamefont {N.~M.~R.}\ \bibnamefont
  {Peres}}, \bibinfo {author} {\bibfnamefont {K.~S.}\ \bibnamefont
  {Novoselov}}, \ and\ \bibinfo {author} {\bibfnamefont {A.~K.}\ \bibnamefont
  {Geim}},\ }\href {\doibase 10.1103/RevModPhys.81.109} {\bibfield  {journal}
  {\bibinfo  {journal} {Reviews of Modern Physics}\ }\textbf {\bibinfo {volume}
  {81}},\ \bibinfo {pages} {109} (\bibinfo {year} {2009})}\BibitemShut
  {NoStop}%
\end{thebibliography}%


%

\end{document}